\documentclass[conference,10pt]{IEEEtran}
\IEEEoverridecommandlockouts
\usepackage{cite}
\usepackage{amsmath,amssymb,amsfonts}
\usepackage{algorithmic}
\usepackage{graphicx}
\usepackage{textcomp}
\usepackage{xcolor}
\usepackage{subcaption}
\usepackage{url}
\usepackage{tikz}
\usetikzlibrary{arrows.meta,calc,colorbrewer,decorations,decorations.pathreplacing,decorations.pathmorphing,fit,matrix,patterns,positioning,shapes,spy}
\usepackage{pgfplots}
\pgfplotsset{compat=1.14}
\usepgfplotslibrary{groupplots,statistics}
\usepackage{standalone}
\usepackage{booktabs}

\begin{document}

\title{ICNLoWPAN -- Named-Data Networking for Low Power IoT Networks}

\author{\IEEEauthorblockN{Cenk G\"undo\u{g}an}
\IEEEauthorblockA{\textit{HAW Hamburg}\\
\small cenk.guendogan@haw-hamburg.de}
\and
\IEEEauthorblockN{Peter Kietzmann}
\IEEEauthorblockA{\textit{HAW Hamburg} \\
\small peter.kietzmann@haw-hamburg.de}
\and
\IEEEauthorblockN{Thomas C. Schmidt}
\IEEEauthorblockA{\textit{HAW Hamburg} \\
\small t.schmidt@haw-hamburg.de}
\and
\IEEEauthorblockN{Matthias W\"ahlisch}
\IEEEauthorblockA{\textit{Freie Universit\"at Berlin} \\
\small m.waehlisch@fu-berlin.de}
}

\maketitle

\begin{abstract}
Information Centric Networking is considered a promising communication technology for the constrained IoT, but 
NDN was designed only for standard network infrastructure.

	In this paper, we design and evaluate an NDN convergence layer for low power lossy links that (1)  augments the NDN stateful forwarding with a highly efficient name eliding,  (2) devises stateless compression schemes for standard NDN use cases, (3) adapts NDN packets to the small MTU size of IEEE 802.15.4, and (4) generates compatibility with 6LoWPAN so that IPv6 and NDN can coexist on the same LoWPAN links.  Our findings indicate that stateful compression can reduce the size of NDN data packets by more than  70~\% in realistic examples. Our experiments show that for common use cases ICNLoWPAN saves 33~\% of transmission resources over NDN, and about 20~\% over 6LoWPAN. 	
\end{abstract}


\section{Introduction} \label{sec:intro}

 The Internet of Things (IoT) has been identified as a potential   
deployment area for Information Centric Networks (ICN) \cite{adiko-sind-12}, since 
infrastructureless access to content, resilient forwarding, and in-network 
data replication demonstrated notable advantages over the
traditional host-to-host approach on the Internet \cite{bmhsw-icnie-14,sblwy-ndnti-16}.  
Named Data Networking (NDN) \cite{jstp-nnc-09} has matured to a prominent, widely implemented flavor of ICN that strictly couples data delivery to consumer requests. 
Recent studies \cite{gklp-ncmcm-18} have shown that NDN outperforms CoAP~\cite{RFC-7252} and MQTT-SN~\cite{mqttsn12}, the corresponding IP-based data services for the constrained IoT. However, NDN lacks mechanisms for packet adaptation and compression to comply with limitations of wireless links prevalent at the low power IoT edge.

Common  edge networks in the constrained IoT are built from low power and
  lossy radios (see "LLN" in \cite{RFC-7228}) such as IEEE 802.15.4 \cite{IEEE-802.15.4-11}, Bluetooth Low Energy (BLE), or LoRA.  Characteristics of LLNs include an unreliable environment, low
 bandwidth transmissions, and increased latencies.  IEEE 802.15.4
 admits a maximum physical layer packet size of 127 octets.
 With 6LoWPAN \cite{RFC-4944,RFC-6282}, the IP-world has created 
  a convergence layer that provides appropriate 
   frame encapsulation formats, packet header compression and link fragmentation for IPv6
   packets in IEEE 802.15.4 networks. The ICN world has not yet developed corresponding features. 

With this paper, we close the gap of LoWPAN convergence for NDN. We leverage its potential of stateful forwarding for eliding names on paths  and design highly efficient compression primitives that outperform 6LoWPAN. Our evaluations reveal significant gains in packet reduction, energy consumption, and reliability. In addition to stateful and stateless compression, we also contribute a fragmentation scheme as well as a framing compatible to the 6LoWPAN techniques. 
Real implementations under RIOT OS~\cite{bghkl-rosos-18} and experimentation on a testbed of current IoT hardware demonstrate the feasibility, robustness and energy efficiency of our approach.

The remainder of this paper is organized as follows. The subsequent Section \ref{sec:rel-work} discusses the problem of NDN LoWPAN adaptation and related work. In Section \ref{sec:icn-lowpan}, we introduce our ICNLoWPAN convergence layer and detail out IPv6-ICN coexistence, on-link fragmentation, and compression. A thorough evaluation of the compression benefits follows in Section \ref{sec:eval}. Finally, we conclude with an outlook in Section~\ref{sec:c+o}. 

\section{Problem Space and Related Work}\label{sec:rel-work}

The Internet of Things inherently connects numerous devices of substantial heterogeneity.
In this work, we focus on deployment use cases that bundle low-end and
battery-operated microprocessors in wireless networks, where packet transmission distinctly dominates power consumption.
The challenges we face in such scenarios are manifold and range from limited MTUs, lossy links and mobility to link layers that lack basic protocol features, such as frame encapsulation formats (c.f.~EtherTypes in Ethernet).

NDN couples name-based routing from TRIAD~\cite{gc-acrsi-01} with stateful forwarding from DONA~\cite{kccek-dona-07} and seamlessly leverages in-network caching on the forwarding plane.
The fundamental request-response semantic on the network layer of NDN requires an \textit{Interest} message and a returning \textit{data} message.
Both message types utilize flexible Type--Length--Value (TLV) header fields to allow for generic and extensible packet formats to the cost of space efficiency.
Name TLVs are essential to NDN and thus always appear in Interest as well as in data messages.
Depending on the naming scheme, human-readable Name TLVs make up the largest part of a request, and also of a response for many IoT use cases .
We explore related work that copes with strict message length limitations using header compression and fragmentation in IP networks first, and then discuss proposed solutions for NDN.

IPv6 mandates a minimum MTU of 1280 bytes for each link and thus precludes a proper IPv6 operation in low power networks with small-sized MTUs.
The IETF designed and extends a set of protocols for constrained IoT deployments where the 6LoWPAN convergence layer is an integral part of.
It is situated below the network layer and provides packet encapsulation, stateless and stateful header compression as well as a protocol independent link fragmentation scheme.
A generic header compression (GHC)~\cite{RFC-7400} extends 6LoWPAN with an LZ77 flavored approach to deal with headers and header-like payloads that are not covered by the 6LoWPAN compression specification.
While 6LoWPAN proves necessary for an interoperable and interconnected host-centric IoT, the same challenges remain open for information-centric IoT deployments.

Shang~\textit{et al.}~\cite{saz-dinps-16} proposed a lightweight link fragmentation scheme that prepends a 3-byte fragmentation header to each NDN fragment to allow for messages larger than the limiting MTU of IEEE~802.15.4.
This custom header further supports a minimal protocol identification by distinguishing normal NDN messages from fragmented messages.
However, this protocol encapsulation collides with the 6LoWPAN encapsulation and thus disregards interoperability with 6LoWPAN, especially in multi-interface and multi-protocol deployments.

Another approach was presented by Mosko~\textit{et al.}~\cite{draft-mosko-icnrg-beginendfragment}, which extends NDN with a new message type that encapsulates each message fragment.
It also adds complexity to the state machine for each peer to initialize fragment sequence numbers and perform corrective actions in case of drifting sequences after packet loss.
A similar approach is the NDN Link Protocol (NDNLP)~\cite{sz-nlpd-12}, which features fragmentation and reassembly as well as ARQ mechanisms.
It provides packet encapsulation to distinguish between normal NDN messages and link acknowledgments.
The last two approaches add overhead in terms of memory consumption and error-control messages which is a disadvantage for low power use cases.

A secure fragmentation for content-centric networks that does not rely on hop-by-hop reassembly and therefore decreases latency was derived by Ghali~\textit{et al.}~\cite{gnotw-sfcn-15}.
Each fragment is securely signed according to NDN semantics and can be cached on intermediate routers.
The authors propose a new ContentFragment message type that includes a Name TLV for forwarding purposes.
Since NDN names are theoretically of unlimited length, duplicating names for each ContentFragment message adds a significant overhead, which naturally is controversial in constrained IoT networks.
Furthermore, sporadic disruptions and mobility in LLNs do not guarantee a successful handover of each individual fragment to complete the reassembly.

Yang~\textit{et al.}~\cite{ys-lntsc-18} focused on bandwidth reduction and an improved storage utilization by translating long names into short names for local communication.
In this regard, a sensor node registers a prefix at a sink node and receives a shorter prefix, e.g., a hash as a name replacement.
Ingress messages that traverse the sink node are updated to include the short name and egress messages respectively are updated to include the long name.
The proposed local name translation is transparent to nodes outside the local IoT network.
However, we argue that a centralized approach to handle name translations with registration and cancellation procedures adds complexity that limits the scalability in large deployments.

In the following section, we concentrate on a fully distributed convergence layer that preserves compatibility with IP LoWPAN, but takes advantage of the information-centric characteristics to obtain a generically applicable, efficient though uncomplex compressive encoding named ICNLoWPAN. 

\section{ICNLoWPAN} \label{sec:icn-lowpan}

\begin{figure}
    \centering
    \input{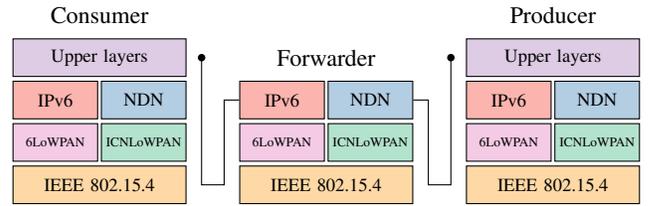}

\begin{tikzpicture}[>=latex]
    \node[traversal-154] (154-a) {IEEE~802.15.4};
    \node[traversal-154,anchor=west,xshift=0.70cm] (154-b) at (154-a.east) {IEEE~802.15.4};
    \node[traversal-154,anchor=west,xshift=0.70cm] (154-c) at (154-b.east) {IEEE~802.15.4};

    \node[traversal-6lowpan,anchor=south west,yshift=0.5mm] (6lowpan-a) at (154-a.north west) {6LoWPAN};
    \node[traversal-6lowpan,anchor=south west,yshift=0.5mm] (6lowpan-b) at (154-b.north west) {6LoWPAN};
    \node[traversal-6lowpan,anchor=south west,yshift=0.5mm] (6lowpan-c) at (154-c.north west) {6LoWPAN};

    \node[traversal-icnlowpan,anchor=south east,yshift=0.5mm] (icnlowpan-a) at (154-a.north east) {ICNLoWPAN};
    \node[traversal-icnlowpan,anchor=south east,yshift=0.5mm] (icnlowpan-b) at (154-b.north east) {ICNLoWPAN};
    \node[traversal-icnlowpan,anchor=south east,yshift=0.5mm] (icnlowpan-c) at (154-c.north east) {ICNLoWPAN};

    \node[traversal-ipv6,anchor=south,yshift=0.5mm] (ipv6-a) at (6lowpan-a.north) {IPv6};
    \node[traversal-ipv6,anchor=south,yshift=0.5mm] (ipv6-b) at (6lowpan-b.north) {IPv6};
    \node[traversal-ipv6,anchor=south,yshift=0.5mm] (ipv6-c) at (6lowpan-c.north) {IPv6};

    \node[traversal-ndn,anchor=south,yshift=0.5mm] (ndn-a) at (icnlowpan-a.north) {NDN};
    \node[traversal-ndn,anchor=south,yshift=0.5mm] (ndn-b) at (icnlowpan-b.north) {NDN};
    \node[traversal-ndn,anchor=south,yshift=0.5mm] (ndn-c) at (icnlowpan-c.north) {NDN};

    \node[traversal-app,anchor=south west,yshift=0.5mm] (app-a) at (ipv6-a.north west) {Upper layers};
    \node[traversal-app,anchor=south west,yshift=0.5mm] (app-c) at (ipv6-c.north west) {Upper layers};

    \draw ([xshift=2mm]app-a.east) node[circle,fill=black,inner sep=1pt] {} |- ([xshift=-2mm]154-b.west)
    |- (ipv6-b.west) (ndn-b.east) -| ([xshift=2mm]154-b.east) -| ([xshift=-2mm]app-c.west) node[circle,fill=black,inner sep=1pt] {};

    \node[yshift=0.75mm,font=\small,anchor=south] at ($(ipv6-b.north west)!0.5!(ndn-b.north east)$) {Forwarder};
    \node[yshift=0.75mm,font=\small,anchor=south] at (app-a.north) {Consumer};
    \node[yshift=0.75mm,font=\small,anchor=south] at (app-c.north) {Producer};
\end{tikzpicture}
    \caption{Stack traversal in a 6LoWPAN and ICNLoWPAN.}
    \label{fig:icnltraversal}
\end{figure}

ICNLoWPAN provides a convergence layer that maps ICN packets onto constrained link layer technologies to enable pure NDN deployments without running as an overlay on top of IP.
Our convergence layer includes features such as link fragmentation, protocol separation on the link layer level as well as stateless and stateful header compression mechanisms.
Fig.~\ref{fig:icnltraversal} shows the overall network stack of a 6LoWPAN and an ICNLoWPAN deployment in parallel for a consumer, a forwarder, and a producer.
Both convergence layers are situated between the link layer and the actual network layer, such that each message traverses them.
This allows for a transparent operation without the need for modifications on the network layer.

\subsection{6LoWPAN Dispatching Framework}
6LoWPAN defines a dispatching framework in~\cite{RFC-4944}, where each frame is prepended with a 1-byte dispatch type and a possible dispatch header.
Several dispatch types exist already, e.g., for stateless IPv6 compression, a mesh header for mesh-under routing purposes, link fragmentation, and extensions to expand the limited universe of possible dispatch types.
One of those extensions is the 1-byte page switch dispatch~\cite{RFC-8025}, which arranges dispatch types into 16 pages and signals a context switch to the packet parser to choose the proper page before interpreting subsequent dispatch types.

ICNLoWPAN integrates into the 6LoWPAN dispatching framework by defining four new dispatch types for Interest and data messages that are either compressed or uncompressed.
Since page 0 and page 1 are already reserved for 6LoWPAN usage, we allocate these dispatches from page 2 to allow for coexistence with 6LoWPAN deployments.
A prepended page switch dispatch before the very first ICNLoWPAN dispatch is thus necessary as an indicator to the dispatch parser.

\begin{figure}[b]
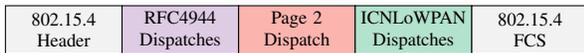

    \centering
    \includestandalone{figures/lowpanfull}
    \caption{IEEE~802.15.4 encapsulated ICNLoWPAN message.}
    \label{fig:icnlmessage}
\end{figure}

A typical ICNLoWPAN message encapsulated in an IEEE~802.15.4 frame is shown in Fig.~\ref{fig:icnlmessage}.
RFC4944 dispatches are optional and may include all dispatch types defined in~\cite{RFC-4944}.
Note the 1-byte page 2 dispatch before the first ICNLoWPAN dispatch.
To switch back to 6LoWPAN dispatches after an ICNLoWPAN dispatch, another page switch dispatch to page 0 or 1 is necessary.

A major benefit of reutilizing the 6LoWPAN dispatching framework is to share a common code base for dispatch handling.
Notably for multi-interface and multi-protocol deployments that make use of IPv6 and NDN simultaneously, using the same code components in resource-constrained devices is exceptionally valuable for minimizing RAM and ROM requirements.

\subsection{Fragmentation}
Reusing the 6LoWPAN dispatching framework enables ICNLoWPAN to seamlessly benefit from the protocol independent link fragmentation scheme defined in \cite{RFC-4944}.
It is thus possible to fragment large NDN messages to fit the limited maximum physical packet sizes of low power link layers, such as 127~bytes for IEEE~802.15.4.

Practically, a fragmented NDN message includes a 4-byte fragmentation dispatch header that lists the original datagram size and a datagram tag to identify fragments of differing packets.
Subsequent fragments further include an additional 1-byte datagram offset of the payload in the dispatch header.
Fragments are reassembled on the next hop and passed to the NDN network stack as typical NDN Interest or data messages.
The 6LoWPAN fragmentation scheme does not define ARQ mechanisms to recover lost fragments,
but rather relies on corrective actions of the link layer.
This allows for implementations with minimal memory footprints.

\subsection{Stateless Compression}
ICNLoWPAN defines a stateless header compression scheme with the main purpose of reducing header overhead of NDN packets.
This is of particular importance for link layers with small MTUs and for increasing energy conservation of battery-operated and wirelessly connected devices.
Corresponding dispatch headers in the ICNLoWPAN packet provide the rule set for decompressing NDN messages before handing the packet over to the NDN network stack.

\subsubsection{TLV Compression}
The NDN header format is solely composed of TLV fields to encode header data.
The advantage of TLVs is its native support of variable-sized data.
The main disadvantage of TLVs is the verbosity that results from storing the \textit{type} and \textit{length} of the encoded data.
Annotating each data requires two bytes extra (\textit{type + length}) for each header field.

The stateless header compression scheme of ICNLoWPAN makes use of compact bit fields to indicate the presence of mandatory and optional TLVs in the uncompressed packet.
The order of set bits in the bit fields corresponds to the order of each TLV in the packet.
Each \textit{type} that is present in the bit field is thus elided from the actual TLV representation, which translates to a reduction from 1 byte to 1 bit.
Further compression is achieved with eliding the \textit{length} of TLVs that either represent fixed-length header data, or where the length can be assumed from surrounding TLVs.
We also achieve smaller encodings by specifying sane default configurations for IoT use cases.

\subsubsection{Name Compression}
A Name TLV is substantial to NDN messages and usually consists of several name components, each of variable size.
An Interest message essentially carries the Name TLV and a data message returns either the same TLV,
or a more specific variant with an equal prefix.
The NDN TLV encoding requires at least two bytes for each name component (\textit{type} + \textit{length}) and an extra two bytes for the outer-most Name TLV.
The TLV overhead for a name is displayed in Eq.~\ref{eq:ndnname}, where $\left| c \right|$ is the number of components.
\begin{equation}
TLV \ overhead = 2 + 2 \cdot \left| c \right|
\label{eq:ndnname}
\end{equation}

\begin{figure}[h]
    \centering
    \input{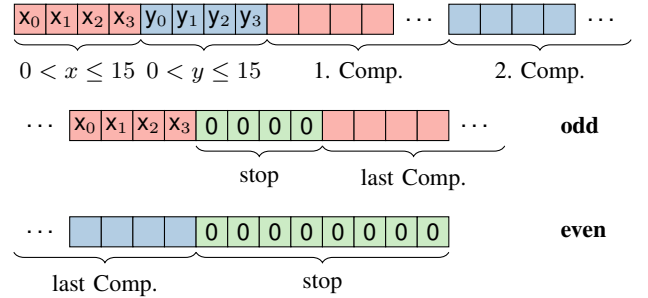}

\begin{tikzpicture}
    \matrix[mbitfull] (m1) {%
        x$_0$ \& x$_1$ \& x$_2$ \& x$_3$ \&%
        y$_0$ \& y$_1$ \& y$_2$ \& y$_3$ \&%
        \& \& \& \& |[draw=none,text width=0.75cm,align=center]| \ldots \&%
        \& \& \& \& |[draw=none,text width=0.75cm,align=center]| \ldots \\%
    };

    \matrix[mbitodd,anchor=north west,yshift=-1.00cm] (m2) at (m1.south west) {%
        |[draw=none,text width=0.75cm,align=center]| \ldots \& x$_0$ \& x$_1$ \& x$_2$ \& x$_3$ \&%
        0 \& 0 \& 0 \& 0 \&%
        \& \& \& \& |[draw=none,text width=0.75cm,align=center]| \ldots \\%
    };

    \matrix[mbiteven,anchor=north west,yshift=-1.00cm] (m3) at (m2.south west) {%
        |[draw=none,text width=0.75cm,align=center]| \ldots \& \& \& \& \&%
        0 \& 0 \& 0 \& 0 \& 0 \& 0 \& 0 \& 0 \\%
    };

    \draw[decorate,decoration={brace,mirror,raise=2pt,amplitude=4pt}] (m1-1-1.south west) -- node[font=\small,below=7pt] {$0 < x \leq 15$} (m1-1-4.south east);
    \draw[decorate,decoration={brace,mirror,raise=2pt,amplitude=4pt}] (m1-1-5.south west) -- node[font=\small,below=7pt] {$0 < y \leq 15$} (m1-1-8.south east);
    \draw[decorate,decoration={brace,mirror,raise=2pt,amplitude=4pt}] (m1-1-9.south west) -- node[font=\small,below=7pt] {1. Comp.} (m1-1-13.south east);
    \draw[decorate,decoration={brace,mirror,raise=2pt,amplitude=4pt}] (m1-1-14.south west) -- node[font=\small,below=7pt] {2. Comp.} (m1-1-18.south east);

    \draw[decorate,decoration={brace,mirror,raise=2pt,amplitude=4pt}] (m2-1-6.south west) -- node[font=\small,below=7pt] {stop} (m2-1-9.south east);
    \draw[decorate,decoration={brace,mirror,raise=2pt,amplitude=4pt}] (m2-1-10.south west) -- node[font=\small,below=7pt] {last Comp.} (m2-1-14.south east);

    \draw[decorate,decoration={brace,mirror,raise=2pt,amplitude=4pt}] (m3-1-1.south west) -- node[font=\small,below=7pt] {last Comp.} (m3-1-5.south east);
    \draw[decorate,decoration={brace,mirror,raise=2pt,amplitude=4pt}] (m3-1-6.south west) -- node[font=\small,below=7pt] {stop} (m3-1-13.south east);

    \node[font=\small\bfseries,xshift=1cm] (odd) at (m2.east) {odd};
    \node[font=\small\bfseries,anchor=west] (odd) at (odd.west |- m3.east) {even};
\end{tikzpicture}
    \caption{Stateless name compression and stop marker for odd and even number of name components.}
    \label{fig:namecomp}
\end{figure}

ICNLoWPAN provides a compression scheme for Name TLVs that drastically reduces the TLV overhead of each nested component.
This compression encodes length fields of two consecutive component TLVs into one byte, using 4 bits each as displayed in Fig.~\ref{fig:namecomp}.
This process limits the length of a component TLV to 15 bytes.
To further elide the outer-most \textit{length} field of the name, this scheme utilizes a stop marker.
For an odd number of components, the stop marker is encoded into the least significant 4 bits of the current \textit{length} byte.
On an even number of components, the full \textit{length} byte is already occupied with two name components.
In this case, a stop byte is appended to the last component as shown in Fig.~\ref{fig:namecomp}.

Compressed names yield a significantly lower TLV overhead as displayed in Eq.~\ref{eq:icnlname}, where $\left| c \right|$ again is the number of components.
The ceil operator handles both cases of odd and even for $\left| c \right|$.

\begin{equation}
TLV \ overhead = \left \lceil{\frac{\left| c \right| + 1}{2}} \right \rceil
\label{eq:icnlname}
\end{equation}

The total TLV overhead reduction for names thus follows from Eq.~\ref{eq:ndnname} and Eq.~\ref{eq:icnlname} and is a function of $\left| c \right|$: $\left \lceil{1.5 \cdot \left| c \right|} \right \rceil + 1$.

\begin{figure*}[h!]
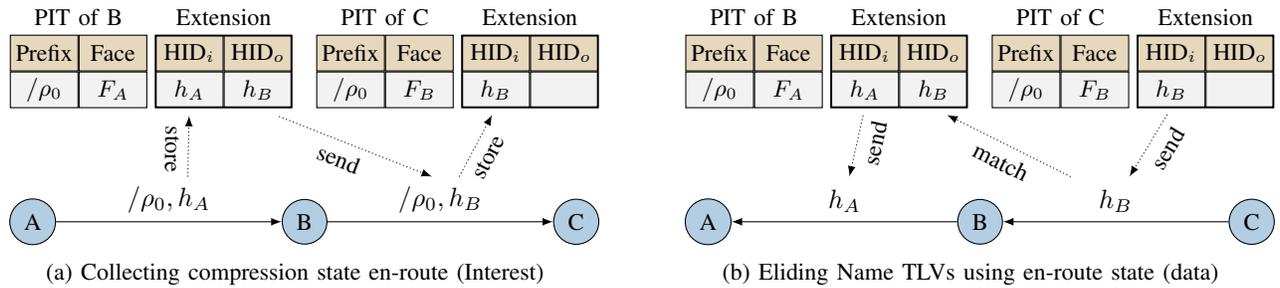

    \begin{subfigure}[c]{\columnwidth}
        \centering
        \input{figures/styles}

\begin{tikzpicture}[>=latex]
    \node[router] (r0) at (0,0) {B};
    \node[router,left=3.00cm of r0] (r1) {A};
    \node[router,right=3.00cm of r0] (r2) {C};
    \draw[->] (r1) to[] node[above] (t1) {$/\rho_0 , h_A$} (r0);
    \draw[->] (r0) to[] node[above] (t2) {$/\rho_0 , h_B$} (r2);

    \matrix[smallpit,inner sep=0pt,yshift=2.0cm,anchor=west] at (r1.west) (pitb1) {%
        Prefix \& Face \& |[splitcell]| \& HID$_{i}$ \& HID$_{o}$\\
        $/\rho_0$ \& $F_A$ \& |[splitcell]| \& $h_A$ \& $h_B$\\
    };
    \node[draw,semithick,inner sep=-0.4pt,fit={(pitb1-1-4.north west)(pitb1-2-5.south east)}] {};
    \matrix[smallpit,inner sep=0pt,yshift=2.0cm,anchor=east] at (r2.east) (pitc1) {%
        Prefix \& Face \& |[splitcell]| \& HID$_{i}$ \& HID$_{o}$\\
        $/\rho_0$ \& $F_B$ \& |[splitcell]| \& $h_B$ \&\\
    };
    \node[draw,semithick,inner sep=-0.4pt,fit={(pitc1-1-4.north west)(pitc1-2-5.south east)}] {};

    \node[font=\small,align=center,anchor=south] at ($(pitb1-1-1.north west)!0.5!(pitb1-1-2.north east)$) {PIT of B};
    \node[font=\small,align=center,anchor=south] at ($(pitb1-1-4.north west)!0.5!(pitb1-1-5.north east)$) {Extension};
    \node[font=\small,align=center,anchor=south] at ($(pitc1-1-1.north west)!0.5!(pitc1-1-2.north east)$) {PIT of C};
    \node[font=\small,align=center,anchor=south] at ($(pitc1-1-4.north west)!0.5!(pitc1-1-5.north east)$) {Extension};

    \draw[->,densely dotted,shorten >= 1mm] (t1.50) -- node[sloped,below,rotate=180,font=\small,pos=0.4] {store} (pitb1-2-4.south);
    \draw[->,densely dotted,shorten <= 3mm] (pitb1-2-5.south) -- node[sloped,below,font=\small,pos=0.5] {send} (t2.110);
    \draw[->,densely dotted,shorten >= 1mm] (t2.50) -- node[sloped,below,font=\small,pos=0.4] {store} (pitc1-2-4.south);
\end{tikzpicture}
        \subcaption{Collecting compression state en-route (Interest)}
        \label{fig:en-route-a}
    \end{subfigure}
    \begin{subfigure}[c]{\columnwidth}
        \centering
        \input{figures/styles}

\begin{tikzpicture}[>=latex]
    \node[router] (r0) at (0,0) {B};
    \node[router,left=3.00cm of r0] (r1) {A};
    \node[router,right=3.00cm of r0] (r2) {C};
    \draw[->] (r2) to[] node[above] (t3) {$h_B$} (r0);
    \draw[->] (r0) to[] node[above] (t4) {$h_A$} (r1);

    \matrix[smallpit,inner sep=0pt,yshift=2.0cm,anchor=west] at (r1.west) (pitb1) {%
        Prefix \& Face \& |[splitcell]| \& HID$_{i}$ \& HID$_{o}$\\
        $/\rho_0$ \& $F_A$ \& |[splitcell]| \& $h_A$ \& $h_B$\\
    };
    \node[draw,semithick,inner sep=-0.4pt,fit={(pitb1-1-4.north west)(pitb1-2-5.south east)}] {};
    \matrix[smallpit,inner sep=0pt,yshift=2.0cm,anchor=east] at (r2.east) (pitc1) {%
        Prefix \& Face \& |[splitcell]| \& HID$_{i}$ \& HID$_{o}$\\
        $/\rho_0$ \& $F_B$ \& |[splitcell]| \& $h_B$ \&\\
    };
    \node[draw,semithick,inner sep=-0.4pt,fit={(pitc1-1-4.north west)(pitc1-2-5.south east)}] {};

    \node[font=\small,align=center,anchor=south] at ($(pitb1-1-1.north west)!0.5!(pitb1-1-2.north east)$) {PIT of B};
    \node[font=\small,align=center,anchor=south] at ($(pitb1-1-4.north west)!0.5!(pitb1-1-5.north east)$) {Extension};
    \node[font=\small,align=center,anchor=south] at ($(pitc1-1-1.north west)!0.5!(pitc1-1-2.north east)$) {PIT of C};
    \node[font=\small,align=center,anchor=south] at ($(pitc1-1-4.north west)!0.5!(pitc1-1-5.north east)$) {Extension};

    \draw[->,densely dotted,shorten <= 1mm] (pitc1-2-4.south) -- node[sloped,below,font=\small,pos=0.45] {send} (t2);
    \draw[->,densely dotted,shorten >= 2mm] (t2) -- node[sloped,below,font=\small,pos=0.45] {match} (pitb1-2-5.south);
    \draw[->,densely dotted,shorten <= 1mm] (pitb1-2-4.south) -- node[sloped,below,font=\small,pos=0.45] {send} (t1);
\end{tikzpicture}
        \subcaption{Eliding Name TLVs using en-route state (data)}
        \label{fig:en-route-b}
    \end{subfigure}
    \caption{Stateful header compression using en-route forwarding state.}
    \label{fig:en-route}
\end{figure*}

\subsection{Stateful Compression}
ICNLoWPAN further employs two stateful compression schemes to enhance size reductions.
These mechanisms rely on shared contexts that are either distributed and maintained in the whole LoWPAN, or are generated and maintained on-demand for a particular Interest-data path.
Our stateless and stateful compressions can be applied in succession to produce huge compression savings, which we will show in our later evaluations.

\subsubsection{LoWPAN-local State}
A context identifier (CID) is a 1-byte number that refers to a particular
conceptual context between networked devices and may be used to replace
frequently appearing information, like name prefixes, suffixes, or meta
information, such as an Interest lifetime.
This allows for a reduction of potentially long data to a single byte.
Shared context has to be initially distributed on compile-time or dynamically maintained on run-time in order for a device to properly encode and decode NDN messages.

When outgoing Interests and data packets traverse the convergence layer,
elements of the message are replaced by one or several CIDs according to a CID translation table.
Context identifiers are appended to the last ICNLoWPAN dispatch and several CIDs may be chained
together by setting the most significant bit of a previous CID.
On reception, the original packet is restored and passed to the network layer.

\subsubsection{En-route State}
In NDN, data messages return Name TLVs of corresponding Interest messages.
Returning names either equal to the original Name TLV, or they contain the original Name TLV as a prefix.
This duplication has huge implications for particularly long names.
We make use of 1-byte HopIDs that replace the full name in returning responses.

While an Interest is forwarded, each hop generates an ephemeral HopID that is tied to an entry of the \textit{Pending Interest Table (PIT)}, which is a fundamental component of NDN and used to match returning responses to open requests on each hop.
HopIDs must be ensured to be unique within the local PIT and to exist only during the lifetime of a PIT entry.
The PIT is extended by two new columns to manage HopIDs. \textit{HID$_i$} for inbound HopIDs and \textit{HID$_o$} for outbound HopIDs.

Before sending an Interest, a hop locally generates a HopID that is stored in the HID$_o$ column.
This Interest then includes the freshly generated HopID along with the name. On the next hop, the HopID is extracted from the Interest and stored in the HID$_i$ column of the respective PIT entry.
The forwarder then generates a new HopID, stores it in the HID$_o$ column of the particular PIT entry, and puts this HopID into the Interest message before it is forwarded to the next hop.
This process is repeated for each hop until the request can be satisfied with the corresponding response as displayed in Fig.~\ref{fig:en-route-a}.

The producer of a returning data message reverses this process by obtaining a HopID from the HID$_i$ column of a PIT entry and encodes it into the response message.
If the returning Name TLV equals the original Name TLV, then the name is fully elided.
Otherwise, the distinct suffix is included along with the HopID.
When a response is forwarded, the contained HopID is extracted and used to match against the correct PIT entry by performing a lookup on the HID$_o$ column.
The HopID is then replaced with the corresponding HopID from the HID$_i$ column before forwarding the response, as visualized in Fig.~\ref{fig:en-route-b}.

\subsection{Name Compression Performance}
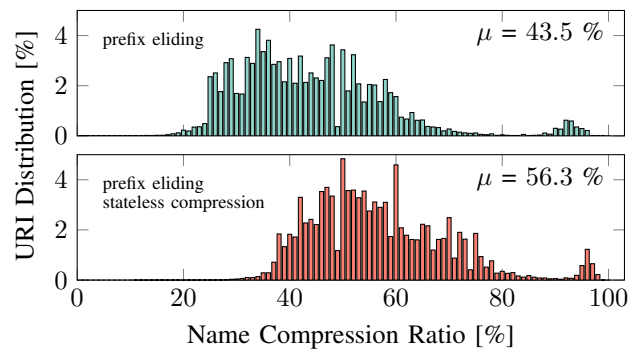
\begin{figure}[b]
    \centering
    \begin{tikzpicture}
    \pgfplotstableread[header=false]{data/namecomp2.csv}{\data}
    \pgfplotstablegetelem{0}{[index]3}\of\data
    \pgfmathsetmacro{\mua}{\pgfplotsretval}
    \pgfplotstablegetelem{0}{[index]4}\of\data
    \pgfmathsetmacro{\mub}{\pgfplotsretval}
    \begin{groupplot}[
            group style={
                group size=1 by 2,
                horizontal sep=0.25cm,
                vertical sep=0.250cm,
            },
            xmin=0, xmax=103,
            ymin=0, ymax=5.0,
            height=3.25cm,
            width=1.00\columnwidth,
            ybar,
            /pgf/bar width=1.5pt,
            xtick align=inside,
            y filter/.code={\pgfmathparse{#1*100}\pgfmathresult},
      ]
      \nextgroupplot[xticklabels={}]
          \addplot[fill=Set3-A] table [x index={0}, y index={1}] {data/namecomp2.csv};
          \node[anchor=north east] at (axis description cs: 0.98,0.98) {$\mu$ = \pgfmathparse{\mua*100}\pgfmathprintnumber[precision=1]{\pgfmathresult} \%};
          \node[anchor=north west,font=\scriptsize] at (axis description cs: 0.03,0.90) {prefix eliding};
      \nextgroupplot[xlabel={Name Compression Ratio [\%]}]
          \addplot[fill=Set3-D] table [x index={0}, y index={2}] {data/namecomp2.csv};
          \node[anchor=north east] at (axis description cs: 0.98,0.98) {$\mu$ = \pgfmathparse{\mub*100}\pgfmathprintnumber[precision=1]{\pgfmathresult} \%};
          \node[anchor=north west,font=\scriptsize] at (axis description cs: 0.03,0.90) {prefix eliding};
          \node[anchor=north west,font=\scriptsize] at (axis description cs: 0.03,0.75) {stateless compression};
    \end{groupplot}
    \path (group c1r1.south west) -- (group c1r2.north west) node[midway,left=4mm,anchor=south,rotate=90] {URI Distribution [\%]};
\end{tikzpicture}
    \caption{Distribution of percental name compression ratios.}
    \label{fig:exp:namecompratio2}
\end{figure}

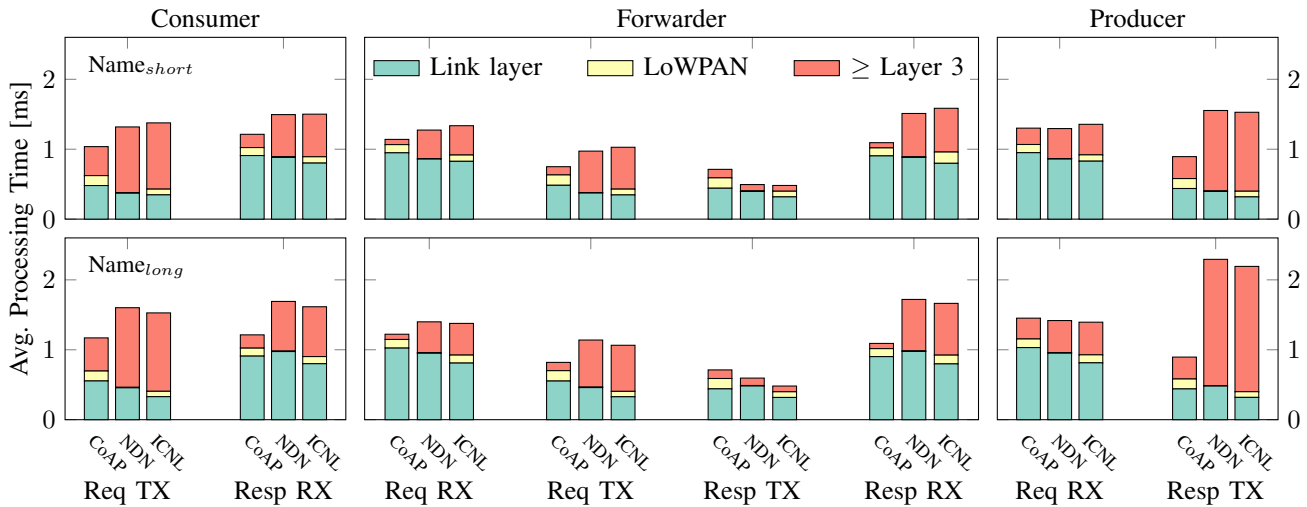
\begin{figure*}
    \centering
    \begin{tikzpicture}
    \begin{groupplot}[
            group style={
                group size=3 by 2,
                horizontal sep=0.25cm,
                vertical sep=0.250cm,
            },
            xmin=0,
            ymin=0, ymax=2.6,
            height=4cm,
            ybar stacked,
            /pgf/bar width=9pt,
            legend cell align=left,
            area legend,
            legend style={draw=none,legend columns=-1,anchor=north,at={(0.5,0.92)},/tikz/every even column/.append style={column sep=15pt},/tikz/every odd column/.append style={column sep=2pt},inner sep=0pt,outer sep=0pt},
            y filter/.code={\pgfmathparse{#1/1000}\pgfmathresult},
            title style={yshift=-2.0mm},
        ]
    \nextgroupplot[
        title=Consumer,
        xmax=9,
        width=0.60\columnwidth,
        xtick={2,7},
        xticklabels={},
    ]
        \addplot[fill=Set3-A] table [y=netif] {data/org-example-temp-1_ALL_4-1-CONSUMER.log.total};
        \addplot[fill=Set3-B] table [y=lowpan] {data/org-example-temp-1_ALL_4-1-CONSUMER.log.total};
        \addplot[fill=Set3-D] table [y=appnet] {data/org-example-temp-1_ALL_4-1-CONSUMER.log.total};
        \node[anchor=north west,font=\small] at (axis description cs: 0.05,0.95) {Name$_{short}$};
    \nextgroupplot[
        title=Forwarder,
        xmax=19,
        width=1.1\columnwidth,
        xtick={2,7,12,17},
        xticklabels={},
        yticklabels={},
    ]
        \addplot[fill=Set3-A] table [y=netif] {data/org-example-temp-1_ALL_4-1-FORWARDER.log.total};
        \addplot[fill=Set3-B] table [y=lowpan] {data/org-example-temp-1_ALL_4-1-FORWARDER.log.total};
        \addplot[fill=Set3-D] table [y=appnet] {data/org-example-temp-1_ALL_4-1-FORWARDER.log.total};
        \legend{Link layer, LoWPAN, $\geq$ Layer 3}
    \nextgroupplot[
        title=Producer,
        xmax=9,
        width=0.60\columnwidth,
        xtick={2,7},
        xticklabels={},
        yticklabel pos=right
    ]
        \addplot[fill=Set3-A] table [y=netif]  {data/org-example-temp-1_ALL_4-1-PRODUCER.log.total};
        \addplot[fill=Set3-B] table [y=lowpan]  {data/org-example-temp-1_ALL_4-1-PRODUCER.log.total};
        \addplot[fill=Set3-D] table [y=appnet]  {data/org-example-temp-1_ALL_4-1-PRODUCER.log.total};
    \nextgroupplot[
        xmax=9,
        width=0.60\columnwidth,
        xtick={2,7},
        xticklabels from table={data/org-example-building-1-floor-4-room-481-temp-8_ALL_4-1-CONSUMER.log.total}{type},
        xticklabel style={ yshift=-0.7cm },
        clip=false,
    ]
        \addplot[fill=Set3-A] table [y=netif] {data/org-example-building-1-floor-4-room-481-temp-8_ALL_4-1-CONSUMER.log.total};
        \addplot[fill=Set3-B] table [y=lowpan] {data/org-example-building-1-floor-4-room-481-temp-8_ALL_4-1-CONSUMER.log.total};
        \addplot[fill=Set3-D] table [y=appnet] {data/org-example-building-1-floor-4-room-481-temp-8_ALL_4-1-CONSUMER.log.total};
        \node[font=\scriptsize,anchor=north west,rotate=-45] at (axis cs:1,0) {CoAP};
        \node[font=\scriptsize,anchor=north west,rotate=-45] at (axis cs:2,0) {NDN};
        \node[font=\scriptsize,anchor=north west,rotate=-45] at (axis cs:3,0) {ICNL};
        \node[font=\scriptsize,anchor=north west,rotate=-45] at (axis cs:6,0) {CoAP};
        \node[font=\scriptsize,anchor=north west,rotate=-45] at (axis cs:7,0) {NDN};
        \node[font=\scriptsize,anchor=north west,rotate=-45] at (axis cs:8,0) {ICNL};
        \node[anchor=north west,font=\small] at (axis description cs: 0.05,0.95) {Name$_{long}$};
    \nextgroupplot[
        xmax=19,
        width=1.1\columnwidth,
        xtick={2,7,12,17},
        xticklabels from table={data/org-example-building-1-floor-4-room-481-temp-8_ALL_4-1-FORWARDER.log.total}{type},
        yticklabels={},
        xticklabel style={ yshift=-0.7cm },
        clip=false,
    ]
        \addplot[fill=Set3-A] table [y=netif] {data/org-example-building-1-floor-4-room-481-temp-8_ALL_4-1-FORWARDER.log.total};
        \addplot[fill=Set3-B] table [y=lowpan] {data/org-example-building-1-floor-4-room-481-temp-8_ALL_4-1-FORWARDER.log.total};
        \addplot[fill=Set3-D] table [y=appnet] {data/org-example-building-1-floor-4-room-481-temp-8_ALL_4-1-FORWARDER.log.total};
        \node[font=\scriptsize,anchor=north west,rotate=-45] at (axis cs:1,0) {CoAP};
        \node[font=\scriptsize,anchor=north west,rotate=-45] at (axis cs:2,0) {NDN};
        \node[font=\scriptsize,anchor=north west,rotate=-45] at (axis cs:3,0) {ICNL};
        \node[font=\scriptsize,anchor=north west,rotate=-45] at (axis cs:6,0) {CoAP};
        \node[font=\scriptsize,anchor=north west,rotate=-45] at (axis cs:7,0) {NDN};
        \node[font=\scriptsize,anchor=north west,rotate=-45] at (axis cs:8,0) {ICNL};
        \node[font=\scriptsize,anchor=north west,rotate=-45] at (axis cs:11,0) {CoAP};
        \node[font=\scriptsize,anchor=north west,rotate=-45] at (axis cs:12,0) {NDN};
        \node[font=\scriptsize,anchor=north west,rotate=-45] at (axis cs:13,0) {ICNL};
        \node[font=\scriptsize,anchor=north west,rotate=-45] at (axis cs:16,0) {CoAP};
        \node[font=\scriptsize,anchor=north west,rotate=-45] at (axis cs:17,0) {NDN};
        \node[font=\scriptsize,anchor=north west,rotate=-45] at (axis cs:18,0) {ICNL};
    \nextgroupplot[
        xmax=9,
        width=0.60\columnwidth,
        xtick={2,7},
        xticklabels from table={data/org-example-building-1-floor-4-room-481-temp-8_ALL_4-1-PRODUCER.log.total}{type},
        yticklabel pos=right,
        xticklabel style={ yshift=-0.7cm },
        clip=false,
    ]
        \addplot[fill=Set3-A] table [y=netif]  {data/org-example-building-1-floor-4-room-481-temp-8_ALL_4-1-PRODUCER.log.total};
        \addplot[fill=Set3-B] table [y=lowpan]  {data/org-example-building-1-floor-4-room-481-temp-8_ALL_4-1-PRODUCER.log.total};
        \addplot[fill=Set3-D] table [y=appnet]  {data/org-example-building-1-floor-4-room-481-temp-8_ALL_4-1-PRODUCER.log.total};
        \node[font=\scriptsize,anchor=north west,rotate=-45] at (axis cs:1,0) {CoAP};
        \node[font=\scriptsize,anchor=north west,rotate=-45] at (axis cs:2,0) {NDN};
        \node[font=\scriptsize,anchor=north west,rotate=-45] at (axis cs:3,0) {ICNL};
        \node[font=\scriptsize,anchor=north west,rotate=-45] at (axis cs:6,0) {CoAP};
        \node[font=\scriptsize,anchor=north west,rotate=-45] at (axis cs:7,0) {NDN};
        \node[font=\scriptsize,anchor=north west,rotate=-45] at (axis cs:8,0) {ICNL};
    \end{groupplot}
    \path (group c1r1.south west) -- (group c1r2.north west) node[midway,left=3mm,anchor=south,rotate=90] {Avg. Processing Time [ms]};
\end{tikzpicture}
    \caption{Intra-stack average processing times for CoAP, NDN, and ICNLoWPAN}
    \label{fig:exp:proctimeall}
\end{figure*}

Our proposed stateless and stateful name compression mechanisms are of simple nature and do not exhaust scarcely available CPU resources.
The compression ratio is highly dependent on 1) the number of name components as they dictate the overall TLV overhead and 2) the prefix length to elide from the name.
As a motivation, we analyze the average name compression performances in Fig.~\ref{fig:exp:namecompratio2} for $356 \cdot 10^6$ real URI paths obtained from the WWW, where each component does not exceed 15~bytes.
Before applying our compression scheme, we encode all names as NDN Name TLVs.
First, we elide the authority component as a prefix for each URI as part of our stateful compression.
This alone yields an average compression ratio of 43.5~\%.
Second, we now apply the stateless compression to reduce TLV overhead and observe a compression ratio of 56.3~\% on average.
We believe that typical names in a low power IoT edge network will yield similar, if not better, compression performances.

\section{Evaluation} \label{sec:eval}

\subsection{Experiment Setup}

\subsubsection{Protocol Comparison}
We consider two different NDN deployments for low power IoT networks.
In the first experiment, we run NDN directly on top of IEEE~802.15.4.
Due to the small MTU of this particular link layer, we limit packet sizes
to a maximum of 100 bytes. In the second experiment, NDN runs on top of
ICNLoWPAN with activated stateless and stateful compression.
We further compare our NDN setups with a typical 6LoWPAN operation
that uses UDP as a transport protocol and CoAP as an application protocol.
We specifically compare against the GET method of CoAP as it provides a request-response pattern analog to NDN.
We deploy our devices in two different network configurations.

\paragraph{Single-hop}
A consumer device has managed FIB entries to a producer device and vice versa.
In our NDN deployment, the producer initially creates all content objects.
The 6LoWPAN producer configures a callback function as a CoAP endpoint for a particular URI to trigger a response.

\paragraph{Multi-hop}
An extra forwarding device is added to the network topology, such that
requests and responses between the consumer and producer traverse through the forwarder.

\subsubsection{Hardware \& Software Platform}
We conduct all our experiments on typical class 2~\cite{RFC-7228} devices that feature
an ARM Cortex-M0+ MCU with 32~kB RAM, 256~kB ROM and up to 48~MHz CPU frequency.
Each device further provides an Atmel AT86RF233~\cite{a-lptzi-09} 2,4~GHz
IEEE~802.15.4 radio transceiver. We set the radio transmission power to 0~dBm, the receiver
sensitivity to -94~dBm and enable the \textit{Smart Receiving} feature, an energy saving mode for idle listening.
For our power consumption measurements we make use of on-board current measurement headers on each of our devices.
We measure currents using a Keithley DMM7510 7\textonehalf~digit graphical sampling multimeter
with 1~MHz sampling rate and control it with external I/O lines to trigger start and stop from events generated by our network stack.
Devices under test are powered by a regulated external DC power supply and connect via UART to a Linux control node to obtain experiment results.

In each experiment, our devices operate RIOT OS version 2018.10. Our NDN deployments use the CCN-lite~\cite{ccn-lite} package and our 6LoWPAN experiments are based on the default GNRC network stack of RIOT OS.
We integrate ICNLoWPAN into the 6LoWPAN module of RIOT OS to reutilize the code base of the dispatching framework and link fragmentation.

\subsubsection{Name Configuration}
Name lengths proportionally affect processing times, packet lengths and consequently energy expenditure during transmissions.
This especially impacts NDN as names are included in requests as well as in responses.
We use two different names in our experiments to measure the effects of our stateless and stateful compression mechanisms.

\paragraph{Name$_{short}$}
We use a short name with 4 components to denote temperature readings produced by a sensor. The name is of the form \texttt{/org/\allowbreak example/\allowbreak temp/\allowbreak $id_x$}, where $id_x$ is an increasing number for each request.
A CID is configured for \texttt{/org}, such that this prefix is elided from all messages.

\paragraph{Name$_{long}$}
We use a long name with 10 components of the form \texttt{/org/\allowbreak example/\allowbreak building/\allowbreak 1/\allowbreak floor/\allowbreak 4/\allowbreak room/\allowbreak 481/\allowbreak temp/\allowbreak $id_x$}.
A CID is configured for the prefix \texttt{/\allowbreak org/\allowbreak example/\allowbreak building/\allowbreak 1/\allowbreak floor/\allowbreak 4/\allowbreak room/\allowbreak 481} to elide a considerable portion of the name.

\subsection{Theoretical Evaluation}
In this first evaluation, we analyze NDN and CoAP packet sizes for a typical IoT scenario using Name$_{long}$ as a CoAP endpoint and as a naming scheme for NDN.
We further configure responses of both protocols to return 4-byte signed integer values that represent temperature sensor readings.

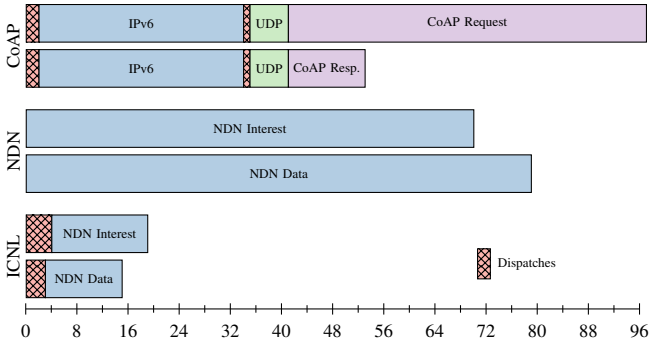
\begin{figure}
    \centering
    \newcommand{\bl}{0.85mm}%

\tikzset {
    encbarpart/.style = { draw, rectangle, anchor=west, minimum height=5.0mm, minimum width=#1, font=\tiny, inner sep=0pt, outer sep=0pt },
    encbar/.style = { matrix of nodes, ampersand replacement=\&, nodes={encbarpart=\bl}, nodes in empty cells, column sep=-\pgflinewidth,anchor=west,inner sep=0pt,outer sep=0pt,font=\tiny },
    disp/.style = { fill=Pastel1-A },
    ip/.style = { fill=Pastel1-B },
    udp/.style = { fill=Pastel1-C },
    coap/.style = { fill=Pastel1-D },
    tl/.style = { fill=Pastel1-E },
    ndn/.style = { fill=Pastel1-B },
    nonce/.style = { fill=Pastel1-I },
    opts/.style = { fill=Pastel1-F },
    content/.style = { fill=Pastel1-G },
    sig/.style = { fill=Pastel1-H },
	coapr/.style = { encbar,
		column 1/.append style = { nodes = { disp } },
		column 2/.append style = { nodes = { ip } },
		column 3/.append style = { nodes = { disp } },
		column 4/.append style = { nodes = { udp } },
		column 5/.append style = { nodes = { coap } },
	},
	ndnint/.style = { encbar,
		column 1/.append style = { nodes = { ndn } },
	},
	ndndata/.style = { encbar,
		column 1/.append style = { nodes = { ndn } },
	},
	icnlint/.style = { encbar,
		column 1/.append style = { nodes = { disp } },
		column 2/.append style = { nodes = { ndn } },
	},
	icnldata/.style = { encbar,
		column 1/.append style = { nodes = { disp } },
		column 2/.append style = { nodes = { ndn } },
	},
}

\begin{tikzpicture}
    \matrix[coapr] (coapreq) at (0cm,3.8cm) {%
        |[encbarpart=2*\bl,postaction={pattern=crosshatch}]| \&%
        |[encbarpart=32*\bl]| IPv6 \&%
		|[encbarpart=1*\bl,postaction={pattern=crosshatch}]| \& |[encbarpart=6*\bl]| UDP \&%
		|[encbarpart=56*\bl]| CoAP Request \\%
    };
    \matrix[coapr] (coapresp) at (0cm,3.2cm) {%
        |[encbarpart=2*\bl,postaction={pattern=crosshatch}]| \&%
        |[encbarpart=32*\bl]| IPv6 \&%
		|[encbarpart=1*\bl,postaction={pattern=crosshatch}]| \& |[encbarpart=6*\bl]| UDP \&%
		|[encbarpart=12*\bl]| CoAP Resp. \\%
    };
    \matrix[ndnint] (ndnint) at (0cm,2.4cm) {%
        |[encbarpart=70*\bl]| NDN Interest \\%
    };
    \matrix[ndndata] (ndndata) at (0cm,1.8cm) {%
        |[encbarpart=79*\bl]| NDN Data \\%
    };
    \matrix[icnlint] (icnlint) at (0cm,1.0cm) {%
        |[encbarpart=4*\bl,postaction={pattern=crosshatch}]| \&%
        |[encbarpart=15*\bl]| NDN Interest \\%
    };
    \matrix[icnldata] (icnldata) at (0cm,0.4cm) {%
        |[encbarpart=3*\bl,postaction={pattern=crosshatch}]| \&%
		|[encbarpart=12*\bl]| NDN Data \\%
    };
    \matrix[encbar] (legend) at (6cm,0.60cm) {%
        |[encbarpart=2*\bl,disp,minimum height=0.4cm,postaction={pattern=crosshatch},label={[xshift=0.1cm]right:Dispatches}]| \\%
    };
	\node[font=\scriptsize, rotate=90, anchor=south,yshift=0.00cm,inner sep=2pt] at ($(coapreq.west)!0.5!(coapresp.west)$) {CoAP};
	\node[font=\scriptsize, rotate=90, anchor=south,yshift=0.00cm,inner sep=2pt] at ($(ndnint.west)!0.5!(ndndata.west)$) {NDN};
	\node[font=\scriptsize, rotate=90, anchor=south,yshift=0.00cm,inner sep=2pt] at ($(icnlint.west)!0.5!(icnldata.west)$) {ICNL};
    \draw (0cm,0cm) -- ++(96*\bl,0cm);
    \foreach \x in {0, 8, ..., 96} {%
        \draw (\x * \bl, 0.075cm) -- (\x * \bl, -0.075cm) node[pos=1, anchor=north,font=\scriptsize] {\x};
    }
    \foreach \x in {4, 12, ..., 96} {%
        \draw (\x * \bl, 0.05cm) -- (\x * \bl, -0.05cm);
    }
\end{tikzpicture}
    \caption{Packet length and structure for different protocols.}
    \label{fig:packetsizes}
\end{figure}

Fig.~\ref{fig:packetsizes} depicts the actual packet sizes for each protocol.
Our CoAP request has a packet size of 97~bytes, where 3~bytes are used for 6LoWPAN dispatches,
32~bytes for the compressed IPv6 header and 6~bytes for the UDP header.
The remaining 56~bytes are used by the actual CoAP message.
The respective CoAP response requires considerably less, which follows from CoAP omitting URIs in responses and using tokens
to match against open requests on the requesting node.
In our setup, CoAP uses 2-byte tokens for each request and returns the exact token in the response.
The NDN Interest message nets to 70~bytes, whereas the returning data message requires 79~bytes.
Contrary to CoAP, returning responses in NDN include the name of the request, or even a more specific and longer name.
The displayed data message further contains empty Signature TLVs.
The equivalent ICNLoWPAN compressed NDN messages are significantly shorter,
where the Interest message reduces down to 19~bytes (72~\% savings) and the data message down to 15~bytes (81~\% savings).
For Interests, this gain is mainly due to leveraging the configured stateful name compression
and data messages naturally benefit from eliding the full returning name.
In addition to the compressed messages, we require a 1-byte page dispatch,
1-byte ICNLoWPAN dispatch, and a 1-byte HopID for Interest and data.
The compressed Interest packet further includes a 1-byte CID indicating the elided prefix for the stateful name compression.

\subsection{Experimental Evaluation}

The theoretical evaluation indicates that ICNLoWPAN substantially
reduces packet sizes in Named-Data IoT networks with pull-driven traffic patterns.
In this experimental evaluation, we want to explore the effects of our
convergence layer on resource-constrained nodes to gauge the feasibility for real-world low power networks.
In particular, we measure
(i) intra-stack processing times,
(ii) average message overhead for a request-response handshake,
(iii) energy expenditure during transmissions for single-hop and multi-hop deployments,
(iv) and effects on reliability when periodically requesting sensor values while we enable randomized and bursty cross-traffic.

\subsubsection{Processing Times}
We first evaluate intra-stack processing times in a multi-hop deployment, where
a consumer requests every 500~ms a specific temperature reading from a producer using Name$_{short}$ and Name$_{long}$.
Timestamps for the link layer, convergence layer (LoWPAN), and upper layers are recorded for each packet using a hardware timer on each device with $\mu$s precision.
The link layer time depicts operations of the RIOT OS radio driver including SPI transfer of the
packet to the radio frame buffer with 5~MHz. This measurement does not contain the actual
transmission or reception of a frame over the wireless medium, as this procedure does not load the CPU.
Time spent in the LoWPAN module includes the handling of dispatch headers and packet (de-)compression.
Processing times for the network layer and beyond either include IPv6, UDP, and CoAP or NDN operations in addition to the actual application on top that issues or satisfies requests.

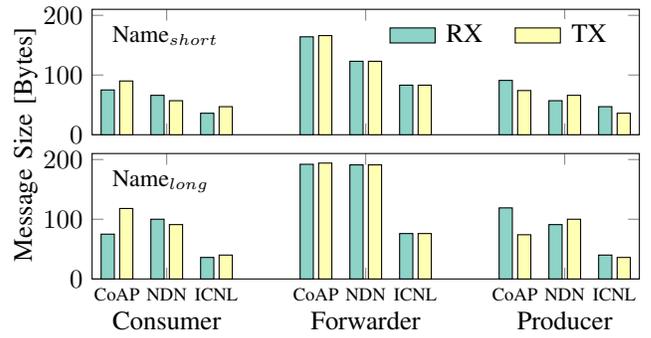
\begin{figure}[t]
    \centering
    \begin{tikzpicture}
    \begin{groupplot}[
            group style={
                group size=1 by 2,
                horizontal sep=0.25cm,
                vertical sep=0.250cm,
            },
            xmin=0, xmax=22,
            ymin=0,ymax=210,
            height=3.25cm,
            width=1.00\columnwidth,
            ybar,
            xtick align=inside,
            /pgf/bar width=5pt,
            legend cell align=left,
            area legend,
            xtick={3,11,19},
            ytick={0,100,200},
            xticklabels={Consumer, Forwarder, Producer},
            xticklabel style = {yshift=-0.30cm},
            legend style={draw=none,legend columns=-1,anchor=north east,at={(0.95,0.92)},/tikz/every even column/.append style={column sep=10pt},/tikz/every odd column/.append style={column sep=2pt},inner sep=0pt,outer sep=0pt},
        ]
        \nextgroupplot[xticklabels={},]
        \addplot[fill=Set3-A, x filter/.expression={\thisrow{type} == 1 ? x : nan}] table [y=bytes] {data/org-example-temp-1_ALL_4-1-ALL.log.stats};
        \addplot[fill=Set3-B, x filter/.expression={\thisrow{type} == 2 ? x : nan}] table [y=bytes] {data/org-example-temp-1_ALL_4-1-ALL.log.stats};
        \legend{RX, TX}
        \node[anchor=north west,font=\small] at (axis description cs: 0.02,0.94) {Name$_{short}$};
    \nextgroupplot[clip=false]
        \addplot[fill=Set3-A, x filter/.expression={\thisrow{type} == 1 ? x : nan}] table [y=bytes] {data/org-example-building-1-floor-4-room-481-temp-8_ALL_4-1-ALL.log.stats};
        \addplot[fill=Set3-B, x filter/.expression={\thisrow{type} == 2 ? x : nan}] table [y=bytes] {data/org-example-building-1-floor-4-room-481-temp-8_ALL_4-1-ALL.log.stats};
        \node[font=\scriptsize,anchor=north] at (axis cs:1,0) {CoAP};
        \node[font=\scriptsize,anchor=north] at (axis cs:3,0) {NDN};
        \node[font=\scriptsize,anchor=north] at (axis cs:5,0) {ICNL};
        \node[font=\scriptsize,anchor=north] at (axis cs:9,0) {CoAP};
        \node[font=\scriptsize,anchor=north] at (axis cs:11,0) {NDN};
        \node[font=\scriptsize,anchor=north] at (axis cs:13,0) {ICNL};
        \node[font=\scriptsize,anchor=north] at (axis cs:17,0) {CoAP};
        \node[font=\scriptsize,anchor=north] at (axis cs:19,0) {NDN};
        \node[font=\scriptsize,anchor=north] at (axis cs:21,0) {ICNL};
        \node[anchor=north west,font=\small] at (axis description cs: 0.02,0.94) {Name$_{long}$};
    \end{groupplot}
    \path (group c1r1.south west) -- (group c1r2.north west) node[midway,left=6mm,anchor=south,rotate=90] {Message Size [Bytes]};
\end{tikzpicture}
    \caption{Average bytes per request--response.}
    \label{fig:exp:bytes}
\end{figure}

Fig.~\ref{fig:exp:proctimeall} displays experiment results for the different roles of a consumer,
forwarder, and producer. We first observe that for the Name$_{short}$ configuration, the additional
processing overhead of ICNLoWPAN does not pay off on the link layer. Thus, some measurements with
ICNLoWPAN take slightly more CPU resources than plain NDN. Conversely, savings on the link layer for our
Name$_{long}$ configuration visibly outperform the ICNLoWPAN processing overhead,
especially for response packets by a decrease of $\approx$~100~$\mu$s per packet.

\subsubsection{Message Sizes}

We now analyze the amount of bytes that are transmitted between consumer, forwarder, and producer when performing a request-response handshake in Fig.~\ref{fig:exp:bytes}.
The captured packets include the message lengths of Fig.~\ref{fig:packetsizes} in addition to 21~bytes for an IEEE~802.15.4 header and 2~bytes for FCS.
While comparing the Name$_{short}$ and Name$_{long}$ configurations, we observe an increase in sent bytes for the CoAP and NDN consumer,
whereas ICNLoWPAN reduces the amount of sent bytes due to a better utilization of our name compression scheme.
We interestingly see that the amount of sent bytes for the CoAP producer stagnates, while the amount for our NDN producer increases.
This follows from the fact that CoAP responses do not include the URI component, but rather a fixed-length token to match against open requests.
In contrast, the NDN data message does include the full name that is obtained from the request, which leads to significantly increased message sizes from Name$_{short}$ to Name$_{long}$.

\subsubsection{Energy Consumption}
\begin{figure*}
    \centering
    \includegraphics{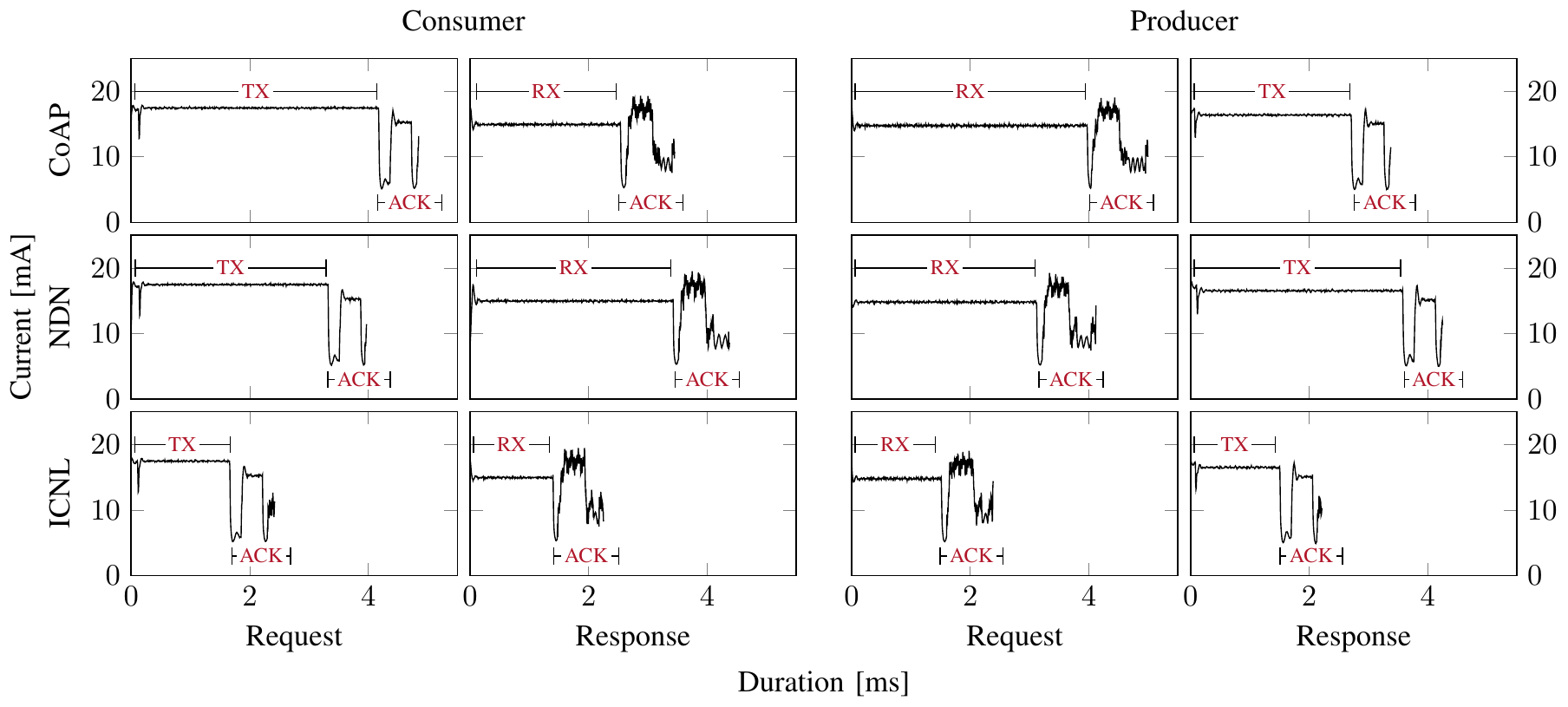}
    \caption{Current consumption for send and receive operations.}
    \label{fig:exp:energy}
\end{figure*}
\begin{table*}
    \renewcommand{\arraystretch}{1.4}
    \centering
    \begin{tabular}{lrrcrrcrr}\toprule
        \phantom{xxxxxxxx} & \multicolumn{2}{c}{Consumer} & \phantom{xxxxxxxx} & \multicolumn{2}{c}{Forwarder} & \phantom{xxxxxxxx} & \multicolumn{2}{c}{Producer}\\
        \cmidrule{2-3} \cmidrule{5-6} \cmidrule{8-9}
        & $Name_{short}$ & $Name_{long}$ & & $Name_{short}$ & $Name_{long}$ & & $Name_{short}$ & $Name_{long}$ \\ \midrule
        CoAP & 548.58 $\mu$J & 612.24 $\mu$J & & 967.41 $\mu$J & 1072.07 $\mu$J & & 464.73 $\mu$J & 517.96 $\mu$J \\
        NDN & 526.23 $\mu$J & 687.26 $\mu$J & & 880.68 $\mu$J & 1152.02 $\mu$J & & 422.55 $\mu$J & 584.82 $\mu$J \\
        ICNL & 466.09 $\mu$J & 487.32 $\mu$J & & 769.17 $\mu$J & 773.97 $\mu$J & & 369.84 $\mu$J & 395.19 $\mu$J \\
        \bottomrule
    \end{tabular}
    \caption{Energy consumption in $\mu$J}
    \label{table:energy}
\end{table*}

ICNLoWPAN decreases the total amount of bytes over the air with both name configurations compared to NDN.
We observe a strong reduction of bytes for responses at the producer for both name configurations, since the name is fully elided.
The reduction is most prominent for the Name$_{long}$ configuration at the forwarder with a drop by 60~\% from 191~bytes down to 76~bytes.
This gain is to be expected, considering that the forwarder is involved in four transmissions.

The packet length during transmission has an immediate effect on the energy consumption.
We measure the current while sending and receiving
messages for each role separately in a single-hop setup and display the results in
Fig.~\ref{fig:exp:energy} for Name$_{long}$. The graphs involve transmission over
the wireless, radio turnaround time as well as link layer frame acknowledgment.
In our setup, sending draws slightly higher current than receiving and the duration of each transmission depends on the packet length.
In fact, the duration of each depicted measurement correlates with the respective message size displayed in Fig.~\ref{fig:packetsizes} and the results showed in Fig.~\ref{fig:exp:bytes},
so that larger messages yield longer periods of operation for sending and receiving.

On the consumer, we observe that our CoAP request requires 5~ms to complete, while the respective NDN request is transmitted in 4~ms, including the reception of acknowledgments for both.
Conversely, the CoAP response is received by the consumer in 3.8~ms, while the NDN response completes in 4.2~ms, including the sending of acknowledgments.
With ICNLoWPAN in operation, we notice a decrease of transmission times by around $\approx$ 50~\% on the consumer due to compressed messages and the resulting shortened media utilization.
As expected, the reduction for responses is more prominent due to fully eliding Name TLVs.
On the producer, we naturally observe mirrored results for each operation.

Given the fact that current draws for transmissions in Fig.~\ref{fig:exp:energy} are mainly similar,
the actual energy consumption is predominated by the transmission durations.
We thus analyze the overall power consumption of a request-response handshake in a multi-hop setup for Name$_{short}$ and Name$_{long}$ including processing times and transmission durations in Tab.~\ref{table:energy}.
Our results indicate an increase in energy usage for each role with the Name$_{long}$ configuration compared to the Name$_{short}$ configuration.
We further notice that our producer spends the least amount of energy, followed by the consumer, and
our forwarder expends nearly double the amount of energy than the producer.
The increased power consumption is inherently consistent with the fact that the forwarder is involved in two request and subsequently in two response transmissions.

The NDN consumer device uses 4~\% less energy for Name$_{short}$ and 12~\% more energy for Name$_{long}$ compared to the CoAP consumer.
This turnaround in energy expenditure for Name$_{long}$ is twofold. 1) NDN has a more verbose name encoding than CoAP and 2) CoAP does not include the URI in the response.
ICNLoWPAN reduces energy usage of our NDN consumer by 11~\% for Name$_{short}$ and 29~\% for Name$_{long}$.
Our NDN producer consumes 9~\% less energy for Name$_{short}$ and 13~\% more energy for Name$_{long}$ compared to our CoAP producer.
The energy consumption reduces by 12~\% for Name$_{short}$ and 32~\% for Name$_{long}$ with an enabled ICNLoWPAN operation compared to NDN.
Since our forwarder interacts with four transmissions, we observe a natural increase in overall expenditures.
The NDN forwarder consumes 9~\% less power for Name$_{short}$ and 7~\% more energy for Name$_{long}$ compared to the CoAP forwarder.
In contrast, ICNLoWPAN reduces the expenditure by 13~\% for Name$_{short}$ and 33~\% for Name$_{long}$.
The trend that ICNLoWPAN yields higher energy savings for Name$_{long}$ becomes apparent.

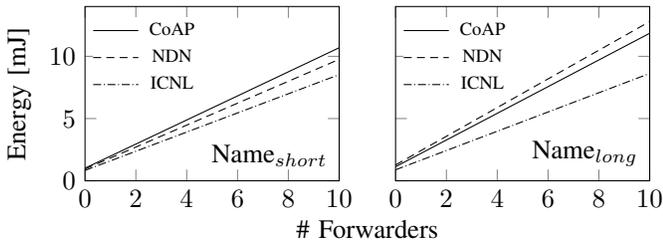
\begin{figure}
    \centering
    \begin{tikzpicture}
    \begin{groupplot}[
            group style={
                group size=2 by 1,
                horizontal sep=0.75cm,
                vertical sep=0.250cm,
            },
            xmin=0, xmax=10,
            ymin=0, ymax=14,
            height=3.9cm,
            width=0.56\columnwidth,
            no markers,
            domain=0:10,
            xtick={0,2,4,6,8,10},
            y filter/.code={\pgfmathparse{#1/1000}\pgfmathresult},
            legend style={draw=none,legend columns=1,anchor=north west,at={(0.03,0.94)},/tikz/every even column/.append style={column sep=15pt},/tikz/every odd column/.append style={column sep=2pt},inner sep=0pt,outer sep=0pt,font=\scriptsize},
        ]
        \nextgroupplot[ylabel={Energy [mJ]}]
            \addplot [solid] { 548.58 + 464.73 + 967.41*x };
            \addplot [densely dashed] { 526.23 + 422.55 + 880.68*x };
            \addplot [densely dashdotted] { 466.09 + 369.84 + 769.17*x };
			\legend{CoAP, NDN, ICNL}
            \node[anchor=south east] at (axis description cs: 0.99,0.03) {Name$_{short}$};
        \nextgroupplot[yticklabels={}]
            \addplot [solid] { 612.24 + 517.96 + 1072.07*x };
            \addplot [densely dashed] { 687.26 + 584.82 + 1152.02*x };
            \addplot [densely dashdotted] { 487.32 + 395.19 + 773.97*x };
			\legend{CoAP, NDN, ICNL}
            \node[anchor=south east] at (axis description cs: 0.99,0.03) {Name$_{long}$};
    \end{groupplot}
    \path (group c1r1.south west) -- (group c2r1.south east) node[midway,below=4mm,anchor=north,rotate=0] {\# Forwarders};
\end{tikzpicture}
    \caption{Total energy consumption for multi-hop networks.}
    \label{fig:exp:energymultihop}
\end{figure}

Finally, we calculate the energy consumption of a full request-response handshake for a multi-hop setup with a varying number of forwarders between a consumer and producer in Fig.~\ref{fig:exp:energymultihop}.
For all setups, we see an increase in expenditure for CoAP, NDN, and ICNLoWPAN with an increasing number of forwarders.
We again notice that the power consumption for NDN surpasses the consumption for CoAP using the Name$_{long}$ configuration due to returning Name TLVs in the response.
ICNLoWPAN clearly reduces the overall energy consumption of an NDN handshake for both configurations Name$_{short}$ and Name$_{long}$.
Interestingly, despite the increase in power consumption for NDN versus CoAP, ICNLoWPAN manages to cut expenditures by $\approx$~25~\% for a setup with ten forwarders compared to NDN.

\subsubsection{Reliability}
\begin{figure}
    \centering
    \includegraphics{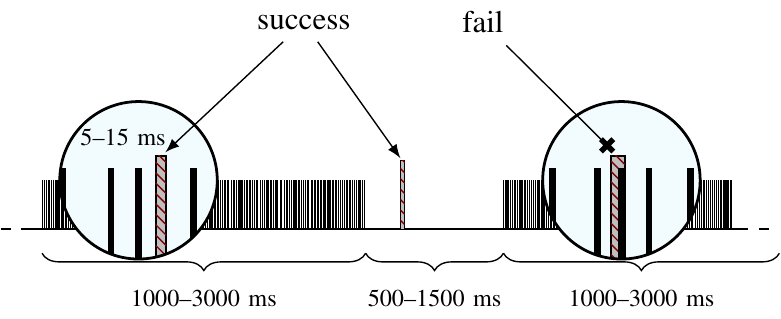}
    \caption{Radio interference due to bursty cross-traffic.}
    \label{fig:exp:crosstraffic}
\end{figure}

\begin{table}[b]
    \renewcommand{\arraystretch}{1.3}
    \centering
    \begin{tabular}{cccccc}\toprule
        & CoAP & \phantom{xxx} & NDN & \phantom{xxx} &  ICNL\\ \midrule
        Producer & 93.53 \% & & 94.05 \% & & 94.40 \% \\
        Consumer & 73.25 \% & & 75.98 \% & & 93.06 \% \\
        \bottomrule
    \end{tabular}
    \caption{Packet Reception Ratio (PRR)}
    \label{table:reliability}
\end{table}
In this experiment we investigate the reliability of a typical data retrieval setup in our single-hop
deployment, where a consumer periodically requests temperature values from a producer every 300~ms.
We additionally generate bursty cross-traffic in randomized intervals to a third party
in order to mimic a dense network with multiple devices that periodically request sensor readings.
Fig.~\ref{fig:exp:crosstraffic} illustrates the configured traffic pattern of our cross-traffic with
each burst consisting of 200 UDP packets in succession with a 5--15~ms delay in between each
transmission and a radio silence interval of 500--1500~ms.
It is worth noting that our cross-traffic only adds wireless interference as the IEEE~802.15.4
frames are not delivered to the devices due to automatic MAC address filtering in hardware of the radio
module.
We further disable CSMA/CA and frame retransmissions of the radio transceiver to explore reliability gains without distortions of automatic corrective actions in hardware.
Indeed, disabling such hardware features in real deployments is not a far-fetched scenario as sophisticated link layer protocols that employ time-slotted algorithms must satisfy strict time constraints,
while CSMA/CA and retransmissions lead to indeterministic and drifting media accesses.
In fact, several radio transceivers that dual operate IEEE~802.15.4 and Bluetooth Low Energy do not even support such mechanisms in hardware, but rather rely on software implementations.

Tab.~\ref{table:reliability} lists the packet reception ratio (PRR) for consumer and producer roles using the Name$_{long}$ configuration.
For each deployment the number of received requests on the producer lies within $\approx$~93--94~\%.
Conversely, the percentage of successfully received responses on each consumer clearly varies between $\approx$~73--76~\% for CoAP and NDN and $\approx$~93~\% for our ICNLoWPAN operation.

The performance gain of ICNLoWPAN results from strongly compressed packets which lead to a significantly reduced on-air time for the low power wireless transmission (see Fig.~\ref{fig:exp:energy}).
This reduces collision probability with interferer traffic, especially when responses are sent during a burst as shown in Fig.~\ref{fig:exp:crosstraffic}.
Furthermore, responses of a producer always follow the successful reception of a request.
Reduced transaction times with ICNLoWPAN leave more time to the next interferer transmission within a burst, which further reduces the probability of overlapping cross-traffic compared to the NDN and CoAP operation.

\section{Conclusions} \label{sec:c+o}

IoT networking has proven to benefit from information-centric architectures in several directions. In this paper, we worked out the components for adapting NDN to a LoWPAN edge:  compression, framing, and fragmentation. By leveraging the NDN stateful forwarding for compression on path, we could again take specific advantage of the information-centric approach in low power lossy IoT networks.   

 Extensive measurements of our implementation on IoT hardware revealed these benefits in comparison with plain NDN and the IP world (CoAP over   6LoWPAN). Our experimental results clearly showed that ICNLoWPAN outperforms NDN and CoAP in terms of media utilization as well as energy consumption. ICNLoWPAN further reduces end-to-end latencies in multi-hop scenarios, and  contributes to an improved reliability in lossy environments while preserving battery resources. Depending on the use case, savings typically range from 20~\% to 33~\%.

With these results, we hope to contribute insights to the community and to encourage deployment of NDN in
the constrained IoT. Our future work will concentrate on extending ICNLoWPAN to different low power
link technologies, such as BLE and align with adaptation layers for wide-area and cellular technologies
to enable LoRA and NB-IoT for a versatile, efficient, and robust information-centric Internet of Things.

\section*{Acknowledgments}
This work was supported in parts by the German Federal Ministry of Research and Education within the project {\em I3: Information-centric Networking for the Industrial Internet}.

\bibliographystyle{IEEEtran}

\end{document}